# Tetrahedrally ferromagnetic correlations and a glassy-freezing anomaly in the breathing pyrochlore magnet $AgInCr_4S_8$ with partial *A*-site disorder

Yuya Haraguchi,[1,*] Masahiro Kitamura,[1] Masaki Gen,[2] Yusuke Nambu,[3] Akira Matsuo,[2] Koichi Kindo,[2] Masaki Kondo,[2] Masashi Tokunaga,[2] and Hiroko Aruga Katori[1]

[1] *Department of Applied Physics and Chemical Engineering, Tokyo University of Agriculture and Technology, Koganei, Tokyo 184-8588, Japan*

[2] *The Institute for Solid State Physics, The University of Tokyo, Kashiwa, Chiba 277-8581, Japan*

[3] *Institute for Integrated Radiation and Nuclear Science, Kyoto University, Kumatori, Osaka, Japan.*

[*] Corresponding author: chiyuya3@go.tuat.ac.jp

We investigate the chromium breathing-pyrochlore sulfide $AgInCr_4S_8$, a chromium-based thiospinel, by synchrotron x-ray and neutron powder diffraction, dc magnetization, and heat capacity. Diffraction confirms the $F\bar{4}3m$ breathing structure with alternating large and small $Cr_4$ tetrahedra, a large breathing ratio ($d'/d$ = 1.106 at 300 K), and substantial Ag/In intermixing on the *A* sublattice (~16%). No structural transition or magnetic Bragg peaks are detected down to 1.5 K. An enlarged low-angle difference plot between the 1.5 and 20 K neutron diffraction patterns shows a weak broad diffuse-like enhancement, consistent with short-range or frozen correlated moments within the sensitivity of the present data. Susceptibility yields a positive Weiss temperature $\theta_W$ = +92 K and a moment enhancement in 30–60 K, while the magnetic entropy released by ~30 K approaches a scale of order $R$ ln 13, together consistent with the development of short-range tetrahedral ferromagnetic correlations and an effective $S$ = 6 cluster-moment picture. A broad susceptibility cusp with ZFC–FC bifurcation and a low-temperature specific-heat anomaly near 9 K indicate a phenomenological glassy-freezing anomaly without long-range order. $AgInCr_4S_8$ provides a benchmark for the interplay of strong breathing distortion and quenched *A*-site disorder in chromium breathing pyrochlores.

## I. INTRODUCTION

Frustrated magnets with a pyrochlore network of spins, where the magnetic ions sit on a three-dimensional array of corner sharing tetrahedra, have provided many examples of unconventional ground states [1,2]. One variant of this geometry that has drawn increasing attention is the so-called breathing pyrochlore lattice, in which the two sets of tetrahedra are not equivalent but alternate in size and exchange strength [3]. In such a breathing pyrochlore (BP) lattice the nearest neighbor exchange along the short bonds on the small tetrahedra, usually called $J$, and that along the long bonds on the large tetrahedra, $J'$, are in general different. This simple bond alternation turns the pyrochlore into a tunable platform: by changing the ratio and even the signs of $J$ and $J'$, one can steer the system through a variety of frustrated magnetic regimes in three dimensions [4].

On the theory side, models on the breathing pyrochlore lattice have been predicted to host a broad range of states, depending on $J$, $J'$, spin size and additional perturbations such as coupling to the lattice [4]. Proposed examples include spin liquid phases and fractionalized excitations in the quantum limit [5-9], superlattice long range orders stabilized by spin lattice coupling [10,11], and topological spin textures such as hedgehog or skyrmion lattices in the presence of competing interactions and spin orbit effects [12,13]. While quantum rare-earth breathing pyrochlores can approach the isolated-tetrahedron limit, here we focus on Cr-based spinel sulfides with $S$ = 3/2 and competing ferro- and antiferromagnetic exchanges.

A prominent materials realization of such Cr-based breathing pyrochlores is provided by *A*-site-ordered spinels of the form $AA'Cr_4X_8$ ($A$ and $A'$: nonmagnetic cations; X = O, S, or Se), where the magnetic moments are $Cr^{3+}$ with $S$ = 3/2 [14]. In these compounds, the contrast in ionic radius and charge between $A$ and $A'$ stabilizes a zinc-blende-type *A*-site order, which imposes a breathing modulation on the Cr pyrochlore network and naturally realizes a three-dimensional breathing lattice of $Cr^{3+}$ spins, as schematically illustrated in Fig. 1. This concept was first demonstrated for the oxides $LiGaCr_4O_8$ and $LiInCr_4O_8$ [15] and subsequently extended to a series of thiospinels $AA'Cr_4S_8$ with $A$ = Li, Cu, Ag and $A'$ = Al, Ga, In [16]. Notably, Haeuseler and Lutz reported a mixed thiospinel with nominal composition $Ag_{0.5}In_{0.5}Cr_2S_4$ in an early vibrational study, but $A$-site ordering and the associated breathing of the Cr pyrochlore sublattice were not resolved [35]. Renewed interest in breathing pyrochlores has since led to detailed crystallographic and physical-property studies across many $AA'Cr_4S_8$ compounds over the past decade [3,14,17-31].

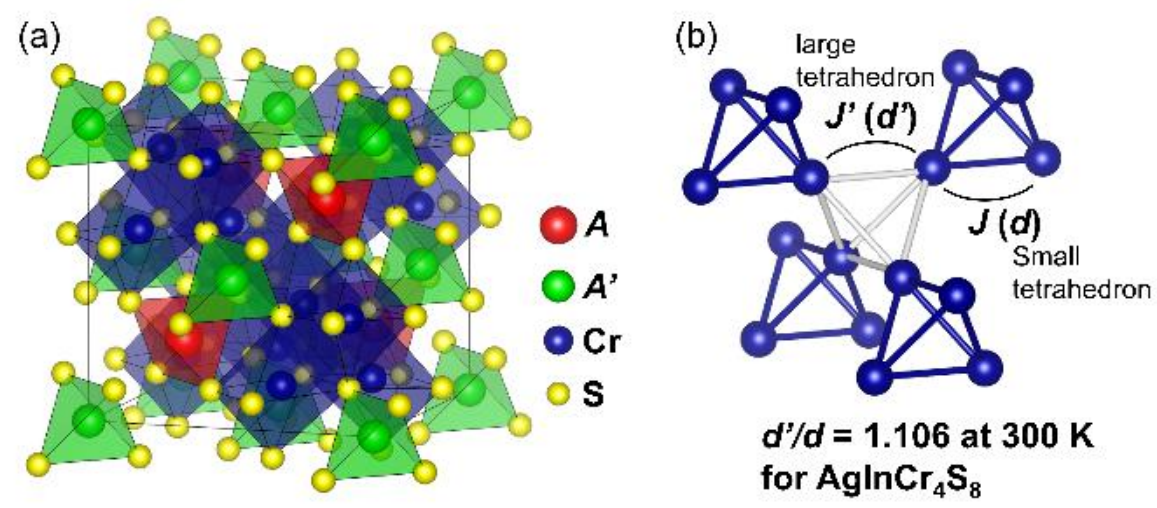

**Figure 1** Schematic illustration of the ideal $A$-site-differentiated chromium spinel $AgInCr_4S_8$ and the corresponding Cr breathing pyrochlore network. (a) Ideal $F\bar{4}3$m framework, in which monovalent $A^+$ and trivalent $A'^{3+}$ cations occupy crystallographically distinct tetrahedral A sites and generate a breathing modulation of the Cr pyrochlore network built from $CrS_6$ octahedra. This panel is a schematic of the ideal $AgInCr_4S_8$ framework; the actual Ag/In intermixing in $AgInCr_4S_8$ is determined by neutron diffraction and discussed in the text. (b) Simplified Cr breathing pyrochlore network consisting of alternating small and large $Cr_4$ tetrahedra. The nearest-neighbor Cr–Cr distances and corresponding exchange interactions are denoted by $d$, $J$ for the small tetrahedra and $d'$, $J'$ for the large tetrahedra, respectively. For $AgInCr_4S_8$, $d'/d$ = 1.106 at 300 K.

A distinctive aspect of chromium breathing pyrochlores is that the nearest neighbor exchanges can take the rather unusual combination of antiferromagnetic $J$ and ferromagnetic $J'$. This arises from the competition between direct Cr–Cr antiferromagnetic exchange on short bonds and ferromagnetic Cr–$X$–Cr superexchange on longer bonds [20]. When the antiferromagnetic $J$ on the small tetrahedra and the ferromagnetic $J'$ on the large tetrahedra are both strong, the four spins on each large tetrahedron can be described by an effective cluster with total spin $S$ = 6, and the problem can be mapped onto a Heisenberg antiferromagnet of effective cluster moments ($S$ = 6) on an effective face-centered-cubic lattice. Experimental evidence of such ferromagnetic clusters has been reported in $CuInCr_4S_8$, where inelastic neutron scattering detects a well-defined cluster excitation and magnetization data are consistent with $S$ = 6 tetrahedral units [29]. Another key ingredient in chromium breathing pyrochlores is strong spin lattice coupling: the exchange constants $J$ and $J'$ depend sensitively on the Cr–Cr distances, so distortions of the lattice can readily relieve frustration. This coupling drives magneto-structural phase transitions at low temperature and under magnetic field [21, 23, 30, 31]. For example, $CuInCr_4S_8$ exhibits a rich field temperature phase diagram including a collinear three up one down phase with a one-half magnetization plateau and a high field pocket [20-24] that has been discussed in terms of a hedgehog or skyrmion lattice [32-34]. More recently, Gen and coworkers have clarified that $CuGaCr_4S_8$ also adopts the $A$-site ordered $F\bar{4}3m$ structure, undergoes a helical magnetic transition at ~ 31 K accompanied by a symmetry lowering, and shows pronounced magnetoelastic and magnetodielectric effects, again highlighting the interplay of cluster magnetism and the lattice [21, 23, 30].

In addition to the balance between $J$ and $J'$ and the magnitude of spin-lattice coupling, quenched disorder on the $A$-sites should be regarded as an additional tuning parameter in Cr-based breathing pyrochlores. $CuAlCr_4S_8$ and $CuGaCr_4S_8$ are close to ideal $A$-site order and develop spin-lattice-coupled helical magnetic order accompanied by symmetry lowering at low temperatures [22,23,30]. By contrast, $CuInCr_4S_8$ exhibits long-range magnetic order characterized by the propagation vector $\boldsymbol{Q}$ = (1, 0, 0), while clear evidence for a zero-field symmetry-lowering lattice distortion has not been established [21]; moreover, both the transition temperature and the field-induced phase diagram can vary between samples [24]. $LiGaCr_4S_8$ contains appreciable Li/Ga site mixing and exhibits glassy (cluster-spin-glass-like) behavior [18,28]; notably, the reported Weiss temperature differs between studies [18,27], again pointing to a sensitivity to quenched randomness. These comparisons suggest that relatively subtle variations in $A$-site disorder can readily perturb the balance among competing exchanges and thereby destabilize coherent long-range order. $LiInCr_4S_8$, typically discussed as having weaker mixing than $LiGaCr_4S_8$, hosts a distinct ordered state [18]. A related fragility is also seen in $LiGa_{1-x}In_xCr_4O_8$, where substitution rapidly suppresses the endmember transitions and yields spin-glass-like freezing for 0.1 ≤ $x \leq 0.6$ [35]. Because breathing ratio, lattice constant, exchange balance, and disorder strength vary simultaneously across compounds, the above trend should be viewed as phenomenological rather than as a single-parameter causality; furthermore, only a limited number of Cr-based breathing-pyrochlore sulfides with substantial *A*-site disorder have been investigated in detail, and a systematic understanding of how breathing distortion and quenched disorder interplay at low temperatures remains limited.

In this work we revisit the Ag–In chromium thiospinel that was previously reported only as a mixed compound with nominal composition $Ag_{0.5}In_{0.5}Cr_2S_4$ based on vibrational spectroscopy, where possible *A*-site ordering

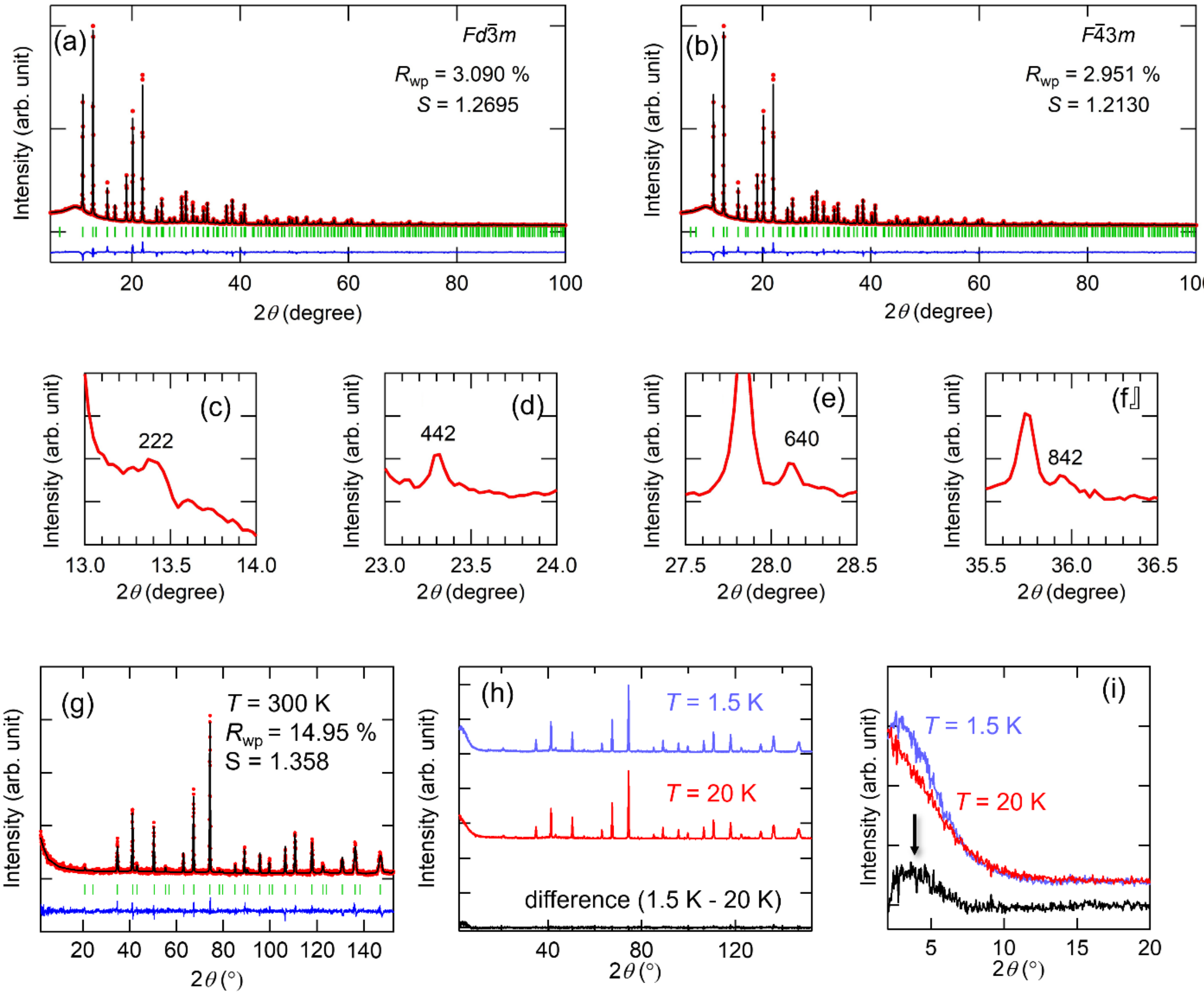

**Figure 2**. Synchrotron x ray and neutron powder diffraction patterns of $AgInCr_4S_8$. Panels (a) and (b) show synchrotron x ray diffraction patterns at 300 K together with Rietveld refinements using the spinel space group $Fd\bar{3}m$ and the breathing pyrochlore space group $F\bar{4}3m$. Red symbols, green tick marks and blue lines represent the observed intensities, Bragg peak positions and calculated profiles, respectively. The $F\bar{4}3m$ model reproduces the data more accurately than $Fd\bar{3}m$. Panels (c)–(f) show enlarged views of XRD peaks that are forbidden in $Fd\bar{3}m$ but allowed in $F\bar{4}3m$. Panel (g) displays the neutron powder diffraction pattern at 300 K and the corresponding Rietveld refinement in $F\bar{4}3m$; red symbols, green tick marks, the black line and the blue line denote the observed intensities, Bragg positions, calculated intensities and the difference between observed and calculated profiles, respectively, confirming that the $F\bar{4}3m$ model also gives an excellent fit to the neutron data. Panel (h) shows neutron diffraction patterns at 1.5 and 20 K and their difference (black line); the absence of any significant difference indicates that no additional Bragg peaks or peak splitting appear down to 1.5 K. Panel (i) shows an enlarged low-angle view of the 1.5 and 20 K neutron diffraction patterns and their difference. The arrow marks a weak broad diffuse-like enhancement in the low-angle difference curve, consistent with short-range or frozen correlated moments. Because of the limited statistics and low-angle background, no quantitative correlation-length analysis is attempted.

**Table 1.** Refined crystallographic parameters of $AgInCr_4S_8$ obtained from Rietveld refinements of synchrotron X-ray powder diffraction (SXRD) data using (a) the disordered spinel model (space group $Fd\bar{3}m$) and (b) the ordered model (space group $F\bar{4}3m$). SOF denotes the site occupancy factor and $B_{iso}$ is the isotropic displacement parameter. Ag and In were treated as a fully disordered mixed site in SXRD; the occupancies were fixed to Ag = 0.5 and In = 0.5 to satisfy the nominal composition.

(a) SXRD, $Fd\bar{3}m$: a=10.23890(6) Å, $S$ = 1.2695, $R_{wp}$ =3.090%

| Atom | Site | $x$ | $y$ | $z$ | SOF | $B_{iso}$ (Å$^2$) |
|---|---|---|---|---|---|---|
| Ag | 8$a$ | 0 | 0 | 0 | 1/2 | 0.962(21) |
| In | 8$a$ | 0 | 0 | 0 | 1/2 | 0.962(21) |
| Cr | 16$d$ | 5/8 | =$x$ | =$x$ | 1 | 0.495(27) |
| S | 32$e$ | 0.38931(10) | =$x$ | =$x$ | 1 | 0.413(30) |

(b) SXRD, $F\bar{4}3m$: a=10.23825(6) Å, $S$ = 1.2130, $R_{wp}$ = 2.951%

| Atom | Site | $x$ | $y$ | $z$ | SOF | $B_{iso}$ (Å$^2$) |
|---|---|---|---|---|---|---|
| Ag1 | 4$a$ | 0 | 0 | 0 | 1/2 | 2.030(89) |
| In1 | 4$a$ | 0 | 0 | 0 | 1/2 | 2.030(89) |
| Ag2 | 4$d$ | 3/4 | 3/4 | 3/4 | 1/2 | 0.619(51) |
| In2 | 4$d$ | 3/4 | 3/4 | 3/4 | 1/2 | 0.619(51) |
| Cr | 16$e$ | 0.36873(20) | =$x$ | =$x$ | 1 | 0.143(37) |
| S1 | 16$e$ | 0.13323(31) | =$x$ | =$x$ | 1 | 0.924(82) |
| S2 | 16$e$ | 0.60751(25) | =$x$ | =$x$ | 1 | 0.598(58) |

**Table 2.** Site occupancies of Ag and In obtained from Rietveld refinement of neutron powder diffraction (NPD) data using the $F\bar{4}3m$ model. The lattice parameter $a$, and the fractional coordinates of Cr and S were fixed to the SXRD-refined $F\bar{4}3m$ values (Table 1b). The Ag/In occupancies were refined with the constraint SOF(Ag) + SOF(In) = 1 at each $A$-site. The "antisite (intermixing) fraction" $x_{mix}$ is defined as the minority occupancy on each $A$-site; thus $x_{mix}$ is approximately 0.16 for both sites in Table 2, and the majority species are In-rich on 4$a$ and Ag-rich on 4$d$.

| Atom | Site | $x$ | $y$ | $z$ | SOF | $B_{iso}$ (Å$^2$) |
|---|---|---|---|---|---|---|
| Ag1 | 4$a$ | 0 | 0 | 0 | 0.16(1) | 0.8(2) |
| In1 | 4$a$ | 0 | 0 | 0 | 0.84(1) | 0.8(2) |
| Ag2 | 4$d$ | 3/4 | 3/4 | 3/4 | 0.84(1) | 0.4(1) |
| In2 | 4$d$ | 3/4 | 3/4 | 3/4 | 0.16(1) | 0.4(1) |
| Cr | 16$e$ | 0.36873(fix) | =$z$ | =$x$ | 1 | 0.14(4) |
| S1 | 16$e$ | 0.13323(fix) | =$z$ | =$x$ | 1 | 0.9(9) |
| S2 | 16$e$ | 0.60751(fix) | =$z$ | =$x$ | 1 | 0.53(16) |

and the associated breathing modulation of the Cr pyrochlore sublattice were not resolved [36]. Motivated by the large ionic radius of $Ag^+$, which can expand the lattice and push the system toward a large-breathing-distortion regime, and with the initial aim of realizing an ideally $A$-site-ordered spinel, we synthesize polycrystalline $AgInCr_4S_8$ and determine its crystal structure by synchrotron x-ray diffraction and neutron powder diffraction. We show that $AgInCr_4S_8$ crystallizes in the noncentrosymmetric space group $F\bar{4}3m$ and realizes a Cr breathing pyrochlore network with the largest breathing ratio ($d'/d$=1.106 at 300 K) among reported Cr-based breathing pyrochlores, while retaining substantial Ag/In intermixing between the two $A$-sites (~16%). Although the $A$-site intermixing was not an intentionally introduced perturbation, it provides a useful opportunity to discuss how quenched disorder and large breathing distortion jointly affect the frustrated magnetism in this family. This establishes $AgInCr_4S_8$ as a concrete platform where $A$-site disorder and large breathing distortion coexist, enabling a systematic assessment of their combined impact on the frustrated magnetism. We then examine its magnetic and dielectric properties by dc magnetization, specific heat, and pulsed-field measurements. The results reveal (i) short-range tetrahedral ferromagnetic correlations on a temperature scale around 30 K and (ii) a phenomenological glassy-freezing anomaly near 9 K, while neutron diffraction down to 1.5 K detects no sharp magnetic Bragg peaks. By comparing $AgInCr_4S_8$ with $A$-site-ordered breathing-pyrochlore magnets such as $CuGaCr_4S_8$ and $CuInCr_4S_8$, we discuss $AgInCr_4S_8$ as an extreme point in the parameter space of breathing distortion and $A$-site disorder and suggest that quenched disorder and bond frustration hinder the helical orders observed in more ordered breathing pyrochlores.

## II. EXPERIMENTAL METHODS

Polycrystalline samples of $AgInCr_4S_8$ were prepared by a conventional solid-state reaction. High-purity powders of Ag (99.9%), $In_2S_3$ (99.99%), Cr (99.9%), and S (99.99%) (all from Kojundo Chemical Lab. Co., Ltd.) were mixed in the stoichiometric ratio, thoroughly ground, and sealed in evacuated quartz tubes. The mixtures were prereacted at 400 °C for 24 h and then fired at 1000 °C for 50 h, followed by furnace cooling to room temperature. The products were reground, pressed into pellets, and annealed again under the same conditions to improve homogeneity. Synchrotron x-ray powder diffraction (SXRD) measurements were carried out at beamline BL-8A of the Photon Factory (KEK, Japan) with a wavelength of λ = 0.6900 Å at 300 K. Neutron powder diffraction (NPD) data were collected at temperatures of 300, 20, and 1.5 K using a fixed-wavelength neutron diffractometer (HERMES, λ =

2.196411 Å) at the JRR-3 research reactor of the Japan Atomic Energy Agency (JAEA), Tokai, Japan.

The diffraction patterns were analyzed by the Rietveld method using the Z-Rietveld program [37] and RIETAN program [38]; structural parameters were refined using the SXRD data, and $A$-site occupancies were refined mainly from the NPD data.

Magnetization measurements were performed using a Quantum Design MPMS in the temperature range 2–300 K and in magnetic fields up to 7 T. Specific heat was measured by the relaxation method using a Quantum Design PPMS between 2 and 300 K. Pulsed-field magnetization and dielectric measurements up to 60 T were carried out at the International MegaGauss Science Laboratory, Institute of Solid State Physics, the University of Tokyo at $T \sim 1.3$–4.2 K [39].

## III. RESULTS

### A. Crystal structure

Figure 2 shows the SXRD pattern of $AgInCr_4S_8$ at 300 K together with Rietveld refinements using the spinel space group $Fd\bar{3}m$ and the breathing pyrochlore space group $F\bar{4}3m$. Several weak reflections forbidden in $Fd\bar{3}m$ but allowed in $F\bar{4}3m$ are reproducibly observed above background (representative examples are shown in the insets of Fig. 2(c)–2(f)), with intensity ratios of 0.1–0.4% relative to the strongest 311 peak. Rietveld fits with the two structural models yield $S = 1.2695$ for $Fd\bar{3}m$ and $S = 1.2130$ for $F\bar{4}3m$, indicating that the $F\bar{4}3m$ model reproduces the SXRD profile slightly better. However, the SXRD fit quality is essentially insensitive to the assumed Ag/In occupancies on the $A$-sites: refinements constrained to Ag = 0.5 and In = 0.5 on each $A$-sites and those forced toward the ideal ordered limit (Ag = 1, In = 0 and vice versa) give nearly the same $S$ values. This reflects the limited x-ray contrast between Ag and In and implies that SXRD alone cannot uniquely determine the degree of $A$-site order versus disorder. As described below, we therefore rely on neutron powder diffraction to quantify the Ag/In occupancies. Importantly, the neutron powder diffraction pattern at 300 K is also well described by the $F\bar{4}3m$ model [Fig. 2(g)], providing the primary confirmation of the $F\bar{4}3m$ breathing average structure. The relatively large $R_{wp}$ value of the neutron refinement should be considered together with the expected reliable factor $R_e = 10.6\%$, which is also large because of the limited counting statistics and background level of the present time-of-flight neutron data. We re-examined the profile parameters, including the Caglioti function, Lorentzian broadening, scale factor, and background, but the $R_{wp}$ value was not significantly improved. Since $S = R_{wp}/R_e$ remains moderate, about 1.3–1.4, the main nuclear Bragg reflections are reasonably reproduced by the $F\bar{4}3m$ model. We therefore use the NPD data primarily to quantify Ag/In occupancies and to test for additional magnetic Bragg peaks, rather than to analyze weak diffuse magnetic scattering quantitatively. We therefore conclude that $AgInCr_4S_8$ adopts the $F\bar{4}3m$ breathing pyrochlore structure at 300 K.

The refined lattice parameter at 300 K is $a$ = 10.23825(6) Å, which is the larger values among reported Cr-based BP sulfides. From the refined Cr and S positions, the nearest-neighbor Cr–Cr distances within the small and large tetrahedra are $d$ = 3.438 Å and $d'$ = 3.801 Å, giving a breathing ratio $d'/d$ = 1.106. This breathing ratio exceeds those reported for $LiACr_4O_8$ ($A'$ = Ga, In) as well as $LiA'Cr_4S_8$ ($A'$ = Ga, In) and $CuA'Cr_4S_8$ ($A'$ = Al, Ga, In), placing $AgInCr_4S_8$ among the most strongly breathing Cr-based BP systems [3, 17-27]. We note that exchange interactions in Cr-based breathing pyrochlores are sensitive to subtle structural changes; therefore, the present statement about a large breathing ratio is strictly tied to the room-temperature average structure determined here.

To quantify the degree of $A$-site cation disorder, we performed Rietveld refinement of the NPD pattern at 300 K using the $F\bar{4}3m$ structure. In this analysis, the positional parameters were fixed to the SXRD values, and only the occupancies of Ag and In on the $A$-sites were refined. The refinement yields $Ag_{0.16(1)}In_{0.84(1)}$ on the $4a$ site and $Ag_{0.84(1)}In_{0.16(1)}$ on the $4d$ site, corresponding to ~ 16 % intermixing between Ag and In. To avoid ambiguity associated with the $A$-site labeling in the SXRD ordered-model table, we adopt the NPD-refined chemical assignment ($4a$: In-rich; $4d$: Ag-rich) and define the intermixing fraction $x$ as the minority occupancy on each $A$-site ($x \approx 0.16$ for both sites). Notably, this In-rich ($4a$) and Ag-rich ($4d$) assignment is opposite to the trend commonly reported for Li- and Cu-based $A$-site-ordered BP sulfides, where the monovalent cation is $4a$-rich and the trivalent cation is 4d-rich. This unusual site preference distinguishes AgInCr4S8 from many previously studied Cr-based BP sulfides. This estimate is model-dependent in the sense that it relies on the average $F\bar{4}3m$ structure with fixed positional parameters; nonetheless, it provides a quantitative measure of substantial $A$-site disorder in $AgInCr_4S_8$ within the present constrained refinement framework. Accordingly, the key conclusion here is the substantial Ag/In intermixing itself; the present data do not require a perfectly unique "ideal" $A$-site order as a reference state.

Figures 2(g) and 2(h) compare NPD patterns at 300, 20, and 1.5 K. Within the experimental resolution, no additional Bragg peaks or peak splitting are observed upon cooling, and the difference pattern between 1.5 and 20 K is essentially featureless. This indicates that no structural phase transition occurs between 1.5 and 300 K, and that no sharp magnetic Bragg peaks are detected down to 1.5 K. Accordingly, we find no evidence for conventional long-range magnetic order in $AgInCr_4S_8$ within the sensitivity of our elastic powder diffraction experiment. To examine possible low-temperature magnetic correlations, we also show an enlarged low-angle view of the 1.5 and 20 K patterns and their difference in Fig. 2(i). The difference curve exhibits a weak broad diffuse-like enhancement at low angles, marked by an arrow, while no sharp magnetic Bragg peak is resolved. This feature is consistent with short-range or frozen correlated moments; however, because of the limited counting statistics and sizable low-angle background, we do not attempt to extract a correlation length or determine a unique spin configuration from the present elastic powder data.

To examine possible low-temperature magnetic correlations, we also plot an enlarged low-angle difference between the 1.5 and 20 K patterns in the inset of Fig. 2(h), after normalization to the nuclear Bragg intensities. The difference curve shows a weak broad diffuse-like enhancement at low angles, while no sharp magnetic Bragg peak is resolved. This feature is consistent with short-range or frozen correlated moments; however, because of the limited counting statistics and sizable low-angle background, we do not attempt to extract a correlation length or determine a unique spin configuration from the present elastic powder data.

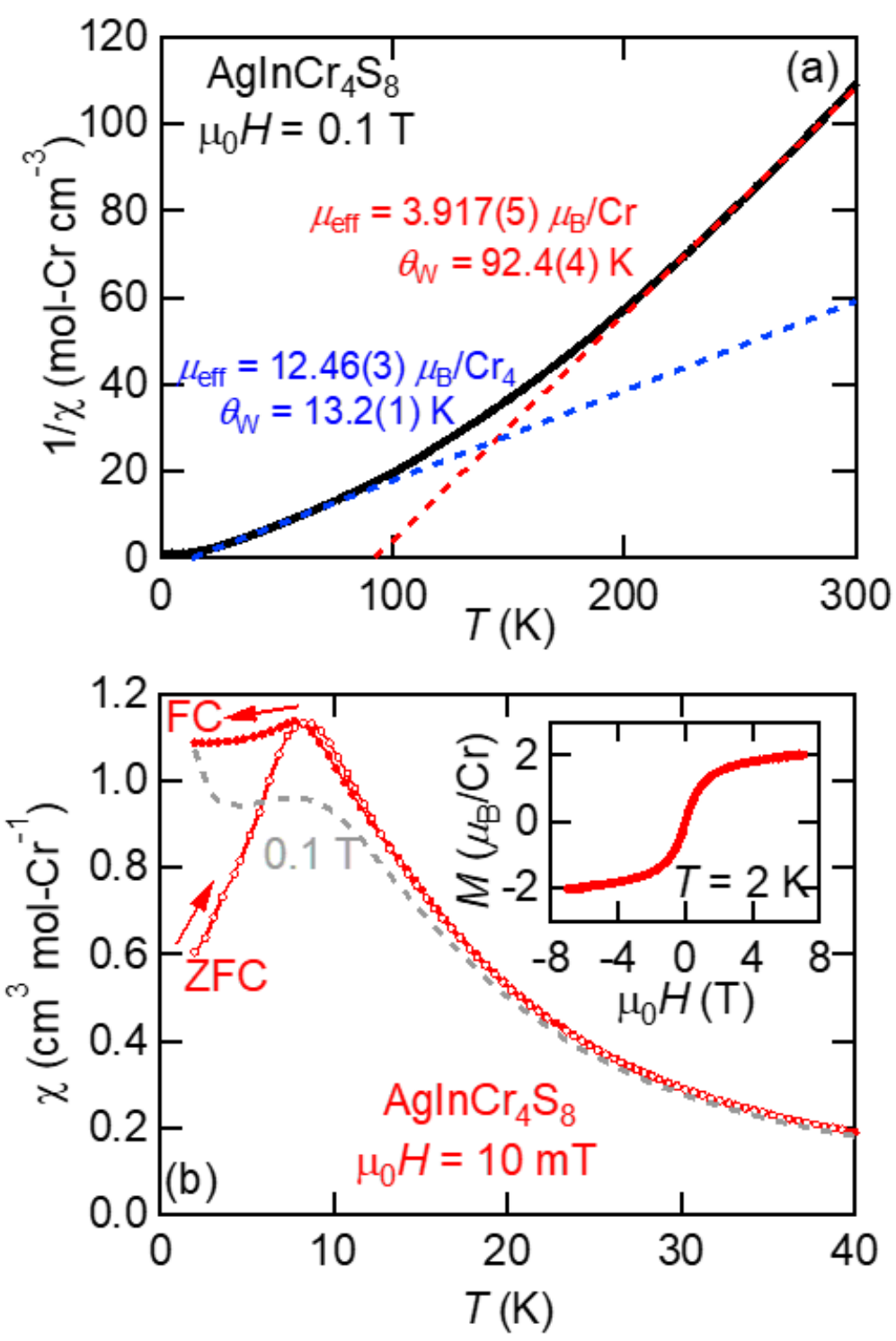

**Figure 3**. Temperature dependence of the magnetic susceptibility of $AgInCr_4S_8$. (a) Inverse susceptibility measured at an applied field of 0.1 T. The red and blue dashed lines show Curie–Weiss fits performed in the high temperature range 200–300 K and in the intermediate range 30–60 K, respectively. (b) Low temperature susceptibility measured at 10 mT under zero field cooled (ZFC) and field cooled (FC) conditions. For comparison, 0.1 T data is also shown. The broad cusp and the bifurcation between the ZFC and FC curves around 9–10 K are used as bulk signatures of a glassy-freezing anomaly. The inset shows the isothermal magnetization at 2 K.

### B. Magnetization

Figure 3(a) displays the temperature dependence of the magnetic susceptibility $\chi(T)$ and its inverse $1/\chi(T)$ measured at $\mu_0 H$ = 0.1 T. The inverse susceptibility follows the Curie–Weiss law between 200 and 300 K; a fit in this range gives an effective moment $\mu_{eff}$ = 3.917(5) $\mu_B$ per Cr and a Weiss temperature $\theta_W$ = +92.4(4) K. The $\mu_{eff}$ value is close to the spin-only value for $Cr^{3+}$ ($S$ = 3/2), indicating localized $Cr^{3+}$ moments, while the large positive $\theta_W$ signifies dominant ferromagnetic interactions. Among Cr-based BP sulfides, this $\theta_W$ is a relatively large positive value.

Upon cooling below ~200 K, $1/\chi(T)$ deviates downward from the high-temperature Curie–Weiss line and becomes approximately linear again between 30 and 60 K. A Curie–Weiss fit in this intermediate range yields an enhanced effective moment $\mu_{eff}$ = 6.23 $\mu_B$ per Cr, corresponding to $\mu_{eff}$ = 12.46 $\mu_B$ per $Cr_4$ tetrahedron. This enhancement is compatible with the growth of short-range ferromagnetic correlations within tetrahedral units, for which an effective cluster-moment description with total spin $S$ = 6 provides a natural heuristic scale. For reference, the spin-only $S$ = 6 cluster is $\mu_{eff} = g\sqrt{(S(S+1))}$ = 12.96 $\mu_B$ for $g$ = 2 (equivalently g = 1.92 for the observed 12.46 $\mu_B$). We therefore regard this behavior as phenomenological evidence of tetrahedral ferromagnetic correlations, rather than as a definitive proof of an isolated-cluster Curie–Weiss regime. Thus, the Curie–Weiss fit in the 30–60 K range should be regarded as phenomenological, and the enhanced moment mainly serves as a marker of a crossover into a correlated (cluster-like) regime. Neutron powder diffraction patterns recorded at 1.5 and 20 K, i.e., below and abov e Tf, show no additional sharp magnetic Bragg peaks [Fig. 2(h)], while the enlarged low-angle difference plot in Fig. 2(i) reveals a weak broad diffuse-like enhancement. Combined with the bulk susceptibility and heat-capacity anomalies, these observations are consistent with a glassy-freezing anomaly of correlated moments. At lower temperatures, $\chi(T)$ exhibits a broad maximum around $T_f \approx 9$ K and a clear bifurcation between zero-field-cooled (ZFC) and field-cooled (FC) curves at low fields [Fig. 3(b)], consistent with a freezing of magnetic degrees of freedom. The small apparent offset between the ZFC and FC curves above $T_f$ is much weaker than the low-temperature bifurcation and is not used as evidence for irreversibility. The low-temperature upturn in the 0.1 T data is mainly associated with the field-cooled response and may reflect field alignment of frozen or slowly fluctuating correlated moments rather than long-range ferromagnetic order. Earlier reports on nominal $Ag_{0.5}In_{0.5}Cr_2S_4$ noted a magnetic anomaly near 14 K without resolving *A*-site ordering/breathing or glassy dynamics [36]; such discrepancies may reflect sample-dependent *A*-site order/intermixing and the associated exchange randomness. We use “spin-glass-like” and “glassy freezing” descriptively for the ZFC–FC bifurcation and broad cusp; dc susceptibility and specific heat alone do not establish a unique universality class or uniquely identify the frozen degrees of freedom. Because frequency-dependent AC susceptibility was not measured in the present study, a dynamic classification of the frozen state remains outside the scope of this work. We therefore use “spin-glass-like” only in a phenomenological sense. AC susceptibility and/or μSR measurements would be required to classify the frozen state more rigorously.

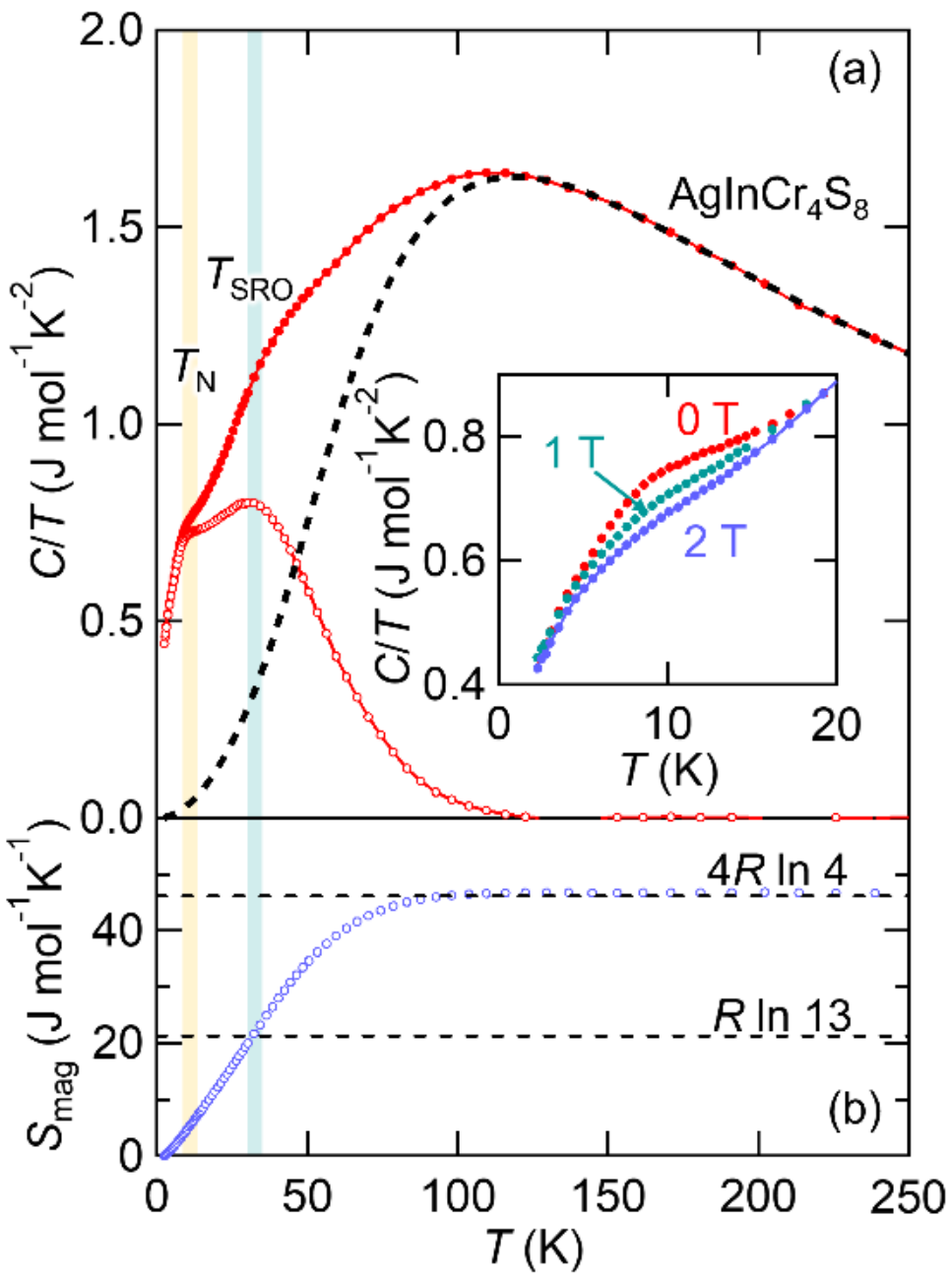

**Figure 4**. Temperature dependence of the specific heat and magnetic entropy of $AgInCr_4S_8$. (a) Specific heat as a function of temperature. Red symbols denote the total specific heat, black symbols the lattice contribution obtained from a Debye–Einstein fit to data above 150 K where the magnetic contribution is negligible compared with the phonon term, and open red symbols the magnetic specific heat estimated as the difference between the total and lattice parts. The inset shows an enlarged view for 0–20 K; red, green and blue symbols correspond to data taken at external fields of 0, 1 and 2 T, respectively. The low-temperature anomaly remains broad under field and does not show a sharp lambda-type feature. (b) Magnetic entropy obtained by integrating the magnetic specific heat divided by temperature. The lattice contribution was calculated from the Debye–Einstein fit and subtracted from the total specific heat to estimate the magnetic part. The horizontal dotted lines indicate reference magnetic-entropy scales, $4R\ln 4$ for four spins with $S$ = 3/2 and $R\ln 13$ as a commonly used reference for an effective $S$ = 6 degree of freedom per formula unit.

### C. Specific heat

Figure 4 shows the temperature dependence of the specific heat divided by temperature, $C/T$, for $AgInCr_4S_8$. To separate the magnetic and lattice contributions, we modeled the lattice specific heat $C_{phonon}(T)$ above 150 K, where the magnetic contribution is negligible, using a Debye term plus two Einstein terms as an empirical interpolation [41]. The best fit was obtained with Debye and Einstein temperatures $\theta_D = 265$ K, $\theta_{E1} = 345$ K, and $\theta_{E2} = 2028$ K and multiplicities $n_1 = 31$ and $n_2 = 8$, respectively. The magnetic specific heat $C_{mag}$ was then obtained by subtracting $C_{phonon}(T)$ from the total specific heat. Given the model dependence of the phonon subtraction, we emphasize the two broad magnetic features and their distinct temperature scales.

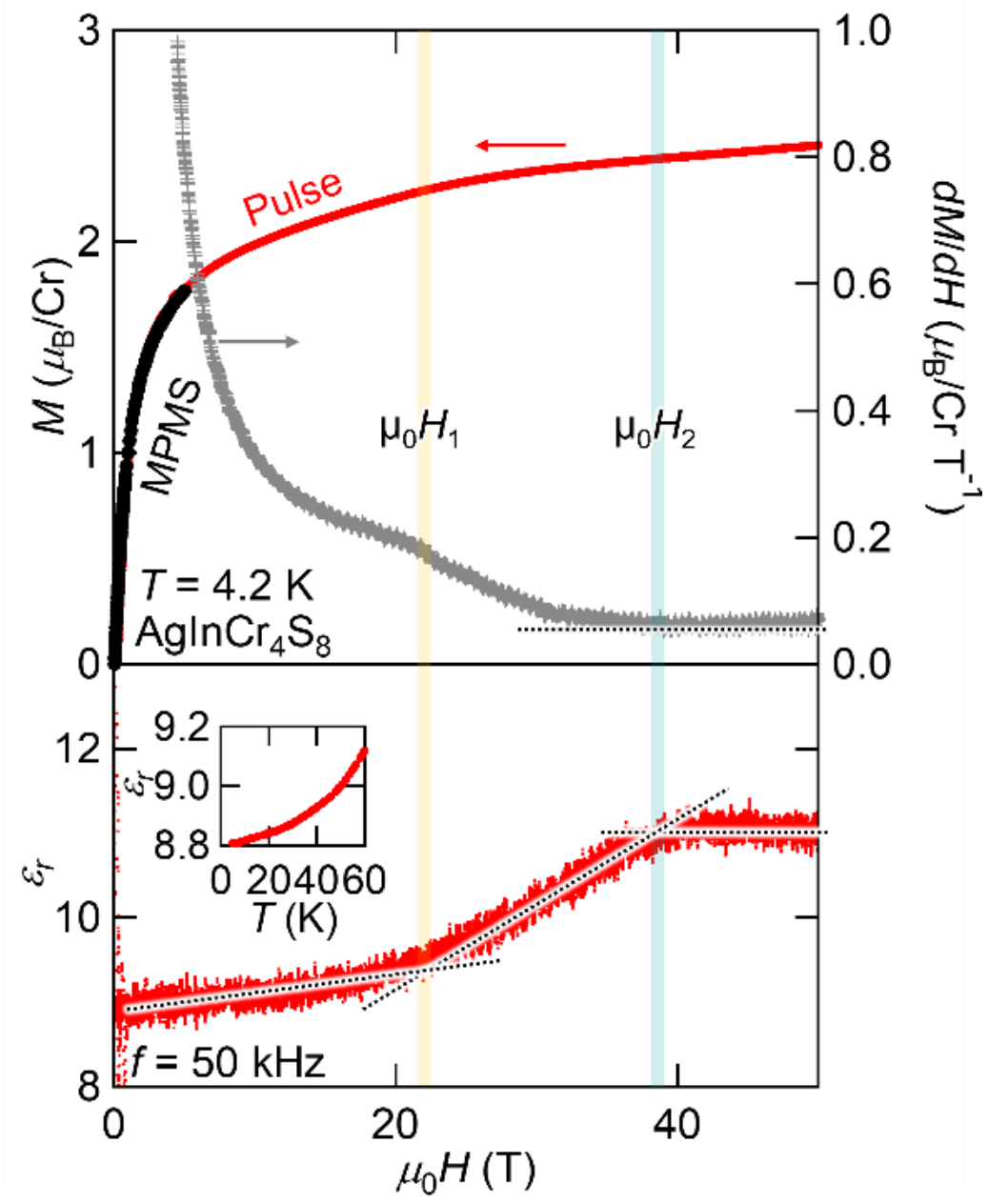

**Figure 5**. High field magnetization and dielectric constant of $AgInCr_4S_8$. Top panel: Isothermal magnetization $M$ as a function of magnetic field and its field derivative $dM/dH$, measured in pulsed fields up to 60 T at 4.2 K. Grey symbols show dc magnetization data taken with an MPMS up to 7 T, which were used to calibrate the pulsed field measurements. Vertical shaded regions mark crossover field scales $\mu_0H_1$ and $\mu_0H_2$ where broad changes in the field evolution are observed. Around $\mu_0H_2$, $M(H)$ shows no sharp kink, but $dM/dH$ becomes nearly field independent. Bottom panel: Relative dielectric constant $\varepsilon_r$ measured in pulsed fields up to 50 T at $T$ = 4.2 K and frequency $f$ = 50 kHz. Dashed lines are guides to eye. $\varepsilon_r$ tends to level off above approximately the same field scale $\mu_0H_2$, indicating coupling between the dielectric response and the field evolution of the correlated magnetic state.

$C_{mag}/T$ exhibits two broad features: a peak around 10 K and a hump centered near $T_{SRO} \approx 30$ K. The lower-temperature peak coincides with the spin-glass-like anomaly in $\chi(T)$, while the higher-temperature hump corresponds to the development of magnetic short-range correlations on the tetrahedral network, consistent with the crossover into the correlated regime inferred from susceptibility. The inset of Fig. 4(a) compares the low-temperature heat capacity under magnetic fields. The anomaly near $T_f$ remains broad and does not develop into a sharp lambda-type peak, consistent with a freezing-like anomaly rather than a conventional long-range ordering transition. Its field dependence is modest in the measured field range, but the broad character of the anomaly is consistent with the gradual suppression or broadening of the ZFC–FC irreversibility under field.

The magnetic entropy $S_{mag}(T)$, obtained by integrating $C_{mag}/T$, is plotted in Fig. 4(b). Because of the broad and overlapping nature of the two anomalies, their entropy contributions cannot be strictly separated; nevertheless, we estimate that $S_{mag} \approx 4.8$ J mol$^{-1}$K$^{-1}$ is released by ~10 K and $S_{mag} \approx 20.0$ J mol$^{-1}$K$^{-1}$ by ~30 K. The latter value is of the same order as $R\ln 13$ = 21.33 J mol$^{-1}$K$^{-1}$ a commonly used reference scale for a single $S = 6$ degree of freedom per formula unit, and is therefore compatible, within the systematic uncertainty associated with the phonon subtraction, with an effective tetrahedral cluster-moment description emerging around $T_{SRO} \approx 30$ K. At higher temperatures, $S_{mag}(T)$ approaches $4R \ln 4$ = 46.08 J mol$^{-1}$K$^{-1}$, consistent with four $S = 3/2$ spins per formula unit and providing an internal consistency check for the lattice subtraction.

### D. High-field magnetization

To probe the high-field magnetization process, we measured the magnetization of $AgInCr_4S_8$ in pulsed magnetic fields up to 50 T at 4.2 K. The upper panel of Fig. 5 shows $M(H)$ together with dc magnetization data taken with an MPMS up to 7 T, which were used to calibrate the absolute scale of the pulsed-field results. Compared with the magnetization processes reported for other Cr-based breathing-pyrochlore sulfides [18,20,23], $AgInCr_4S_8$ exhibits a markedly steep rise of $M(H)$ already in low fields, highlighting an unusually strong net ferromagnetic response consistent with its large positive Weiss temperature. At 50 T, M reaches approximately 2.6 $\mu_B$ per Cr, i.e., about 87% of the fully polarized value of 3 $\mu_B$ per Cr expected for $g = 2$ and S = 3/2. With increasing field, $M(H)$ becomes progressively sublinear, and $dM/dH$ decreases smoothly. A broad change in curvature is visible around $\mu_0H_1$ about 20 T (first shaded region), suggesting that the applied field gradually suppresses an antiferromagnetic component that opposes full polarization. In a tetrahedral-cluster picture, the spins within each $Cr_4$ tetrahedron are largely ferromagnetically correlated, while residual antiferromagnetic couplings between tetrahedra can keep the correlated moments weakly canted at low fields; their progressive field alignment provides a natural qualitative explanation for the smooth crossover around $\mu_0 H_1$. Above $\mu_0 H_1$, the magnetization continues to increase but with a reduced slope. Around $\mu_0 H_2$ about 40 T, $M(H)$ remains smooth without a distinct kink; however, $dM/dH$ approaches an almost field-independent small value above $\mu_0 H_2$, indicating entry into a high-field regime where the magnetization is nearly linear with weak field dependence. Thus, $\mu_0 H_2$ should not be regarded as a sharp anomaly in $M(H)$, but rather as a crossover scale at which the field evolution of the magnetization becomes substantially weaker. Notably, as discussed below, $\varepsilon_r(H)$ also tends to level off above the same field scale, supporting the view that $\mu_0 H_2$ separates two field-evolution regimes of the correlated state. Given possible magnetocaloric and other non-equilibrium effects intrinsic to low-temperature pulsed-field measurements, we treat $\mu_0 H_1$ and $\mu_0 H_2$ as crossover field scales rather than definitive thermodynamic phase boundaries.

### E. Dielectric properties

To probe the dielectric response, we first examine the temperature dependence of the relative dielectric constant $\varepsilon_r$ at zero magnetic field (inset of Fig. 5). Within our resolution, $\varepsilon_r$ shows no sharp anomaly at around 10 K where the spin-glass-like freezing is observed in the dc susceptibility. This contrasts with $A$-site-ordered breathing-pyrochlore magnets such as $CuGaCr_4S_8$, where step-like anomalies in $\varepsilon_r$ have been reported near the magnetostructural transition associated with helical magnetic order [23,24]. The absence of a clear zero-field dielectric anomaly in $AgInCr_4S_8$ is consistent with the lack of a symmetry-lowering structural transition and with the absence of magnetic Bragg peaks in elastic neutron diffraction.

The lower panel of Fig. 5 displays $\varepsilon_r$ measured in pulsed magnetic fields up to 50 T at 4.2 K and at 50 kHz. With increasing field, $\varepsilon_r$ grows gradually and shows its strongest change between about 10 and 40 T, and then tends to level off above $\mu_0 H_2 \sim 40$ T. Importantly, while the isothermal magnetization $M(H)$ itself does not show a sharp anomaly at $\mu_0 H_2$, the derivative $dM/dH$ becomes nearly field independent in the same field range. The dielectric response appears more sensitive to the field evolution of the correlated state than $M(H)$ itself, possibly because it couples to subtle magnetoelastic changes. The simultaneous leveling-off of $dM/dH$ and $\varepsilon_r$ above $\mu_0 H_2$ therefore suggests that $\mu_0 H_2$ marks a crossover between two distinct field-induced regimes of the low-temperature correlated state, even if the transition is broadened into a crossover by disorder and the limited resolution intrinsic to pulsed-field measurements.

Given possible magnetocaloric and other non-equilibrium effects intrinsic to low-temperature pulsed-field measurements, we do not assign a definitive thermodynamic phase boundary from the present dielectric data alone and instead regard the features near $\mu_0 H_2$ as a crossover signature. Nevertheless, the coincidence of the characteristic field scale extracted independently from $dM/dH$ and from $\varepsilon_r(H)$ provides consistent evidence that the spin configuration of the correlated moments couples to the lattice under high magnetic fields. Further probes, such as magnetostriction and field-dependent diffraction measurements, would be valuable to test whether a symmetry change or a superstructure accompanies the crossover near $\mu_0 H_2$.

## IV. DISCUSSION

Our structural and thermodynamic results are consistent with the following picture for $AgInCr_4S_8$: (i) a crossover into a regime of short-range tetrahedral ferromagnetic correlations on a temperature scale around $T_{SRO}$ approximately 30 K and (ii) a glassy-freezing anomaly inferred from bulk signatures at $T_f$ approximately 9 K, without evidence for conventional long-range magnetic order in elastic powder diffraction down to 1.5 K. The crossover in $\mu_{eff}$, the broad entropy release on the 30 K scale, and the absence of sharp magnetic Bragg peaks are mutually consistent with an effective correlated-moment (cluster-like) picture, but we stress that the present dataset does not uniquely determine the microscopic identity or spatial distribution of the frozen entities The enlarged low-angle difference pattern in Fig. 2(i) shows a weak broad diffuse-like enhancement, which is consistent with short-range or frozen correlated moments, but the present neutron data do not allow a quantitative determination of the correlation length or spin configuration. Our "spin-glass-like" terminology is descriptive and based on dc susceptibility, specific heat, and elastic powder diffraction, and we do not assign a unique universality class from these probes alone. A more rigorous dynamic classification would require frequency-dependent AC susceptibility and/or μSR measurements.

First, we consider the exchange interactions in $AgInCr_4S_8$. As shown in Fig. 6, the empirical correlation between Cr–Cr distance and Weiss temperature in chromium spinel chalcogenides suggests that Cr–Cr bonds shorter than about 3.53 Å tend to support antiferromagnetic direct exchange, whereas longer bonds favor ferromagnetic superexchange through chalcogen ions [2,3,42]. In $AgInCr_4S_8$, the small-tetrahedron distance $d$ = 3.438 Å lies below this threshold, while the large-tetrahedron distance $d'$ about 3.801 Å is well above it, implying that the nearest-neighbor exchange $J$ is likely antiferromagnetic and $J'$ likely ferromagnetic. We note, however, that first-principles studies on related BP sulfides indicate that distance-based arguments for $J$ and $J'$ are only semi-quantitative [14,43]. In particular, further-neighbor interactions, $J_2$, $J_{3a}$, and $J_{3b}$, are suggested to be predominantly antiferromagnetic in DFT-based discussions of Cr spinels and breathing-pyrochlore sulfides [14,43], providing an additional source of frustration among tetrahedrally correlated moments. Moreover, the antiferromagnetic contribution arising from direct Cr–Cr exchange is expected to be highly sensitive to the local Cr–Cr bonding geometry (bond direction and local distortions that control orbital overlap). In $AgInCr_4S_8$, quenched lattice disorder associated with Ag/In intermixing can therefore plausibly reduce the effective antiferromagnetic contribution to the nearest-neighbor exchanges $J$ and $J'$. This is consistent with the expectation that the antiferromagnetic direct Cr–Cr exchange on the nearest-neighbor bonds is highly sensitive to local distortions that modify orbital overlap. We do not attempt to quantify how such disorder affects further-neighbor interactions in the present work. Therefore, the ferromagnetic Cr–$X$–Cr superexchange component [43] can become relatively more influential, providing a natural qualitative route to the large positive $\theta_W$ and the unusually strong ferromagnetic response inferred from the low-field magnetization. With these caveats in mind, the positive $\theta_W$ can be understood as reflecting dominant ferromagnetic contributions from $J'$ on the large tetrahedra, partially compensated by antiferromagnetic further-neighbor couplings. In this qualitative picture, the breathing distortion favors strong ferromagnetic correlations on the large tetrahedra, which can be represented by an effective tetrahedral correlated moment (of order $S = 6$ in a minimal cluster description), while the competition among inter-tetrahedral couplings can hinder the establishment of long-range order.

Second, we discuss the influence of $A$-site cation disorder. A key feature of $AgInCr_4S_8$ is the substantial Ag/In intermixing ($x \sim 0.16$), in contrast to Cu-based BP sulfides that are close to the ideally $A$-site-ordered limit. In $CuAlCr_4S_8$ and $CuGaCr_4S_8$, helical magnetic order is accompanied by a symmetry-lowering lattice response, highlighting the role of spin-lattice coupling in stabilizing coherent ordered states [22,23,30]. By contrast, $CuInCr_4S_8$ develops long-range magnetic order characterized by the propagation vector $\boldsymbol{Q}$ = (1, 0, 0); however, clear evidence for a zero-field symmetry-lowering structural distortion has not been established [21], and parts of its transition behavior and field-induced phase diagram can be sensitive to sample quality [24]. On the more disordered side, $LiGaCr_4S_8$ shows appreciable Li/Ga site mixing and is reported to exhibit a cluster-spin-glass-like ground state [18,27,28]. Against this backdrop, $AgInCr_4S_8$ combines an exceptionally large breathing ratio with substantial $A$-site disorder and exhibits a glassy-freezing anomaly rather than coherent long-range order. While multiple material parameters vary across compounds (breathing ratio, lattice constant, exchange balance, and disorder), these comparisons are consistent with the view that quenched $A$-site disorder and the resulting random-bond landscape can readily destabilize long-range ordering tendencies and favor glassy freezing of correlated moments in Cr-based breathing pyrochlores. We emphasize that this is a phenomenological correlation rather than evidence for single-parameter causality. Finally, the unusual In-rich (4*a*) and Ag-rich (4*d*) site preference suggests that the energetic gain from perfect charge ordering may be reduced in this sulfide and may compete with local size- and bonding-related preferences.

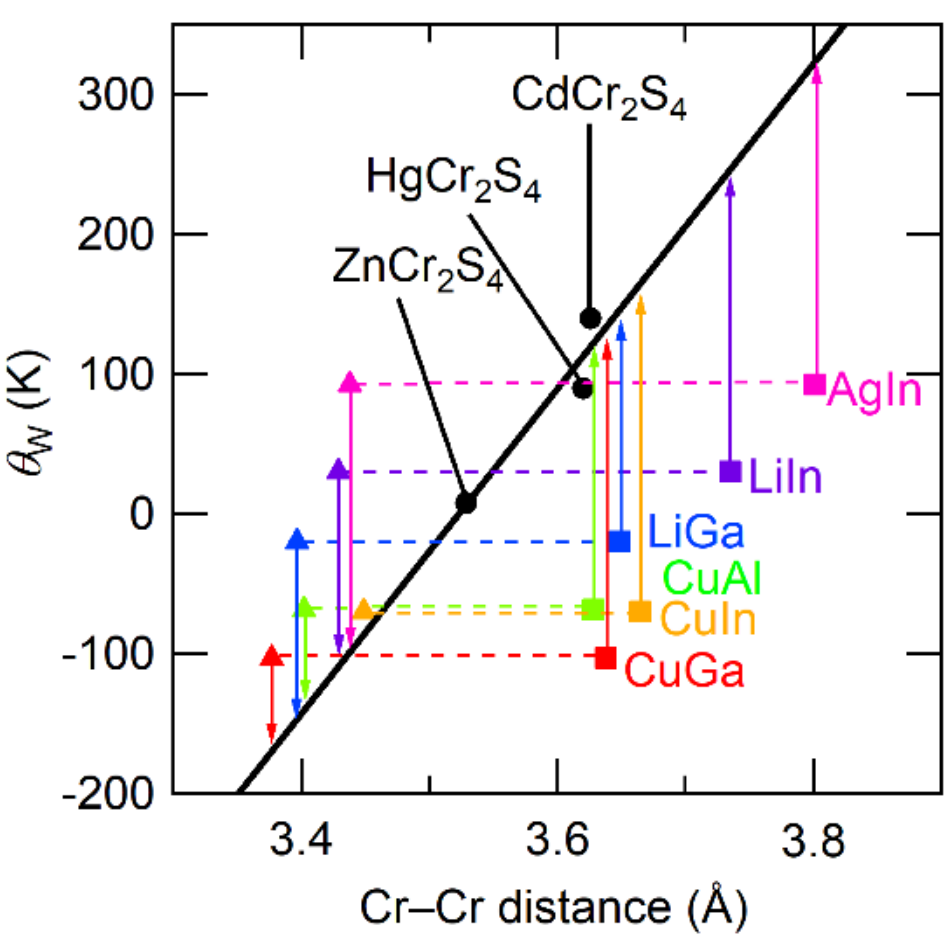

**Figure 6** Relationship between Cr–Cr distance and Weiss temperature for chromium spinel and breathing pyrochlore sulfides. Triangles and squares represent the nearest neighbor Cr–Cr distances $d$ and $d'$ on the small and large tetrahedra of breathing pyrochlores, respectively, while filled circles show the Cr–Cr distances in normal spinel sulfides. The solid line is reproduced from Ref. 18, and the data point for $AgInCr_4S_8$ has been added to this plot. The two Cr–Cr distances of $AgInCr_4S_8$ fall on opposite sides of the crossover around 3.5 Å, suggesting that the nearest neighbor exchange $J$ on the small tetrahedra is predominantly antiferromagnetic whereas $J'$ on the large tetrahedra is ferromagnetic.

Microscopically, random occupancy of $Ag^+$ and $In^{3+}$ on both $A$-sites is expected to modulate Cr–S–$A$–S–Cr superexchange pathways as well as local bond angles and crystal-field environments around $Cr^{3+}$ ions. The resulting spatial variation of inter-tetrahedral couplings (including $J'$ and further-neighbor terms) can introduce random-bond and random-field tendencies for the correlated moments, plausibly hindering the establishment of a unique propagation vector and long-range phase coherence of a spiral state even if the average interactions would otherwise be compatible with it [1]. In this sense, $AgInCr_4S_8$ can be viewed as a candidate for a cluster-like glassy state on the breathing pyrochlore lattice; however, the present dataset does not uniquely determine whether the dominant fluctuating/freezing entities correspond to tetrahedral correlated moments or to larger correlated regions.

It is instructive to contrast this glassy-freezing scenario with canonical spin glasses [38]. In canonical metallic spin glasses, such as dilute alloys with random-sign exchange, the freezing typically involves individual moments, and the freezing temperature tracks the interaction scale. In $AgInCr_4S_8$, by contrast, we observe two well-separated temperature scales: a crossover into a correlated regime around $T_{SRO} \approx 30$ K and a freezing anomaly at $T_f \sim 9$ K with a much smaller additional entropy change. Together with the absence of magnetic Bragg peaks and the presence of a ZFC–FC bifurcation, these bulk results are consistent with a picture in which correlated moments develop first and subsequently undergo glassy freezing. At the same time, the present dataset does not uniquely determine whether the frozen entities correspond to tetrahedral correlated moments or to larger correlated regions. In addition, the field-dependent dielectric response observed here indicates that the correlated state is not purely magnetic but couples to the lattice, which is less typical of canonical metallic spin glasses and may be more naturally discussed in terms of a cluster-glass-like freezing with magnetoelastic degrees of freedom. Because AC susceptibility was not measured, the present work does not determine the frequency-dependent dynamics or the universality class of the frozen state.

Within the family of Cr-based breathing-pyrochlore sulfides studied so far, $AgInCr_4S_8$ combines an exceptionally large breathing ratio ($d'/d = 1.106$ at 300 K) with substantial Ag/In intermixing (~16%). This places $AgInCr_4S_8$ in a regime where strong tetrahedral correlations coexist with quenched $A$-site disorder. More broadly, a pronounced fragility of long-range order against quenched disorder has also been documented in the breathing-pyrochlore oxides $LiGa_{1-x}In_xCr_4O_8$ [35]. Although the microscopic origin of randomness differs (Ga/In substitution rather than Ag/In antisite mixing, and predominantly antiferromagnetic exchanges in the oxides), this phase diagram underscores a general fragility of long-range order against quenched disorder in breathing-pyrochlore magnets. Compared with more $A$-site-ordered BP sulfides that undergo helical magnetic order accompanied by symmetry lowering, $AgInCr_4S_8$ instead exhibits a phenomenological glassy-freezing anomaly without long-range order, consistent with the idea that quenched disorder and bond frustration can hinder the development of coherent helical order. $AgInCr_4S_8$ thus provides a useful benchmark for discussing how breathing distortion, correlated moments, and $A$-site disorder interplay in three-dimensional frustrated lattices.

Finally, we note limitations of the present work. The crossover features in $dM/dH$ and the corresponding dielectric response around $\mu_0H_2 \approx 40$ T indicate a field-driven reorganization within the low-temperature correlated state. However, the smoothness of $M(H)$ and possible non-equilibrium effects in low-temperature pulsed-field measurements prevent us from assigning a sharp thermodynamic phase boundary from the present data alone. Accordingly, we describe these high-field and dielectric signatures in terms of correlated crossover field scales rather than a definitive thermodynamic transition.

Note added. —During the revision of this manuscript, we became aware of a recently accepted study on single-crystalline $AgInCr_4S_8$, which reports incommensurate helical long-range magnetic order below 9.6 K. This behavior differs from the glassy-freezing anomaly observed in the present polycrystalline sample with substantial Ag/In intermixing. Although a direct comparison requires detailed information on the degree of A-site order, stoichiometry, and sample history, the contrast between the two studies further highlights the sensitivity of $AgInCr_4S_8$ magnetism to quenched disorder and local structural details [44].

## V. SUMMARY

In summary, we synthesized $AgInCr_4S_8$ and established by synchrotron x-ray and neutron powder diffraction that it adopts the noncentrosymmetric breathing-pyrochlore spinel structure ($F\bar{4}3m$) with a strongly size-modulated Cr pyrochlore network ($d$ = 3.438 Å, $d'$ = 3.801 Å; $d'/d$ = 1.106) and substantial Ag/In intermixing on the $A$ sublattice (~16% minority occupancy on each $A$-site in our constrained average-structure refinement), while no structural transition or sharp magnetic Bragg peaks are detected down to 1.5 K. An enlarged low-angle neutron-diffraction difference plot [Fig. 2(i)] shows a weak broad diffuse-like enhancement, consistent with short-range or frozen correlated moments, although the present data do not allow a quantitative correlation-length analysis. Magnetization and specific heat further indicate a crossover into a correlated regime around $T_{SRO} \approx 30$ K, consistent with the development of short-range tetrahedral ferromagnetic correlations that can be captured by an effective correlated-moment (cluster-like) description without uniquely proving an isolated-cluster Curie–Weiss regime. At lower temperatures, a broad susceptibility cusp with ZFC–FC bifurcation near $T_f \approx 9$ K and a concomitant low-$T$ $C_{mag}$ anomaly indicate a phenomenological glassy-freezing anomaly without long-range order; the data support freezing of correlated moments but do not uniquely identify the microscopic freezing entities. Overall, $AgInCr_4S_8$ provides a useful benchmark for discussing how strong breathing distortion and quenched $A$-site disorder correlate with frustration-driven glassy-freezing behavior in Cr-based breathing pyrochlores.

## ACKNOWLEDGEMENTS

This work was supported by JST PRESTO Grant Number JPMJPR23Q8 (Creation of Future Materials by Expanding Materials Exploration Space), JST FOREST (JPMJFR202V) and JSPS KAKENHI Grant Numbers. JP25K01496 (Scientific Research (B)), JP25H01649, JP22H05145 and JP23H04616 (Transformative Research Areas (A) "Supra-ceramics"), JP24H01613 (Transformative Research Areas (A) "1000-Tesla Chemical Catastrophe"), JP25H01403 (Transformative Research Areas (B) "Multiply Programmed Layers"), JP22K14002 (Young Scientific Research), and JP24K06953 (Scientific Research (C)). Part of this work was carried out by joint research in the Institute for Solid State Physics, the University of Tokyo (Project Numbers 202311-MCBXG-0021, 202311-MCBXG-0025, 202406-MCBXG-0100, 202406-GNBXX-0095, 202406-MCBXG-0100, 202406-MCBXG-0101, 202411-MCBXG-0033, and 202411-MCBXG-0034). The SXRD experiments were performed with the approval of the Photon Factory Program Advisory Committee (Proposal No. 2022G551).